\documentclass[a4paper,fleqn,usenatbib]{mnras}

\usepackage{mathptmx}

\usepackage[T1]{fontenc}
\usepackage{ae,aecompl}
\pdfoutput=1

\usepackage{graphicx}	
\usepackage{amsmath}	
\usepackage{amssymb}	

\title[Photometry of 9 TNOs and Centaurs]{Photometric observations of nine Transneptunian objects and Centaurs}

\author[T. Hromakina et al.]{
T. Hromakina,$^{1}$\thanks{E-mail: hromakina@astron.kharkov.ua}
D. Perna,$^{2,3}$
I. Belskaya,$^{1}$
E. Dotto,$^{2}$
A. Rossi,$^{4}$
and F. Bisi$^{2,5}$
\\
$^{1}$Institute of Astronomy, Kharkiv V.N. Karazin National University, Sumska Str. 35, Kharkiv 61022, Ukraine\\
$^{2}$INAF -- Osservatorio Astronomico di Roma, Via Frascati 33, I-00078 Monte Porzio Catone (Roma), Italy\\
$^{3}$LESIA -- Observatoire de Paris, PSL Research University, CNRS, Sorbonne Universit{\'e}s, UPMC Univ. Paris 06, Univ. Paris Diderot,\\
Sorbonne Paris Cit{\'e}, 5 place Jules Janssen, 92195 Meudon, France\\
$^{4}$IFAC-CNR, via Madonna del Piano 10, I-50019 Sesto Fiorentino (Firenze), Italy\\
$^{5}$Dipartimento di Fisica, Universit{\`a} di Roma Tor Vergata, via della Ricerca Scientifica 1, 00133 Roma, Italy\\
}

\date{Accepted XXX. Received YYY; in original form ZZZ}

\pubyear{2017}

\begin{document}
\label{firstpage}
\pagerange{\pageref{firstpage}--\pageref{lastpage}}
\maketitle

\begin{abstract}

We present the results of photometric observations of six Transneptunian objects
and three Centaurs, estimations of their rotational periods and corresponding
amplitudes. For six of them we present also lower limits of density values. All
observations were made using 3.6-m TNG telescope (La Palma, Spain). For four
objects -- (148975) 2001 XA255, (281371) 2008 FC76, (315898) 2008 QD4, and 2008
CT190 -- the estimation of short-term variability was made for the first time.
We confirm rotation period values for two objects: (55636) 2002 TX300 and
(202421) 2005 UQ513, and improve the precision of previously reported
rotational period values for other three -- (120178) 2003 OP32, (145452) 2005 RN43,
(444030) 2004 NT33 -- by using both our and literature data. We also discuss
here that small distant bodies, similarly to asteroids in the Main belt, tend
to have double-peaked rotational periods caused by the elongated shape rather
than surface albedo variations.  \end{abstract}




\section{Introduction}

The study of short-term photometric variability of small Solar system bodies
let us to estimate such important physical characteristics as rotation period,
shape and surface heterogeneity. It is believed that rotational rates and
shapes of minor Solar system objects might be a function of their sizes and
densities \citep{Sheppard2008}. The rotational properties of the smallest
objects are thought to be significantly altered since the formation, while the
mid-sized objects are considered to be affected only by quite recent
collisional events, and, finally, the largest among minor bodies population
have their momentum values preserved in the most pristine condition
\citep{Farinella1996, Davis1997, Morbidell2004, Sheppard2008}.

Up to date, more than 2500 Transneptunian objects (TNOs) and Centaurs are
discovered, however, the rotational variability has been measured for less than
100 objects \citep{Harris2016}. We note that only for about 10 of them the
rotation period was determined precisely (code 3 in the A.~Harris database).
Moreover, in most cases two possible values of rotational period are given
because of no confident distinction between single and double-peaked
lightcurves. Such situation can be explained by faintness of these distant
objects, and, as a result, inability for obtaining accurate photometry with
small telescopes. The use of moderate and large telescopes for the purpose of
rotation period determination is usually very limited in time. But to measure a
confident rotational period the object should be observed during at least 2-3
successive nights.

In this paper we present new photometric observations of a selected sample of 9
outer Solar system objects, 6 TNOs and 3 Centaurs. Rotational variability of
four of these objects was observed for the first time. We describe
observational circumstances and data reduction technique and present our
results and their analysis together with the literature data when they are
available. 

\section{Observations and data reduction}

Photometric observations were carried out during two observational runs in March 
and August 2009 at a 3.6-m TNG telescope (La Palma, Spain). We used the DOLORES 
(Device Optimized for the LOw RESolution) instrument equipped with a E2V $4240\times2048$ pixel 
thinned back-illuminated, deep-depleted, Astro-BB coated CCD with a pixel size of 13.5~$\mu$m. 
All  photometric measurements were taken in the broadband \textit{R} filter. 

Data reduction was made following the standard procedure, which included bias 
subtraction from the raw data and flat-field correction, using {\scshape midas} 
software package. We performed only differential photometry. To minimize random 
errors only bright field stars (typically three of them per image) were used. 
The accuracy of photometry measurements is about 0.02-0.03~mag.

In Table~\ref{Table1} we present observational circumstances which include the 
mean UT, heliocentric (r) and geocentric ($\Delta$) distances, and solar phase angle ($\alpha$).

\begin{table*}
	\centering
	\caption{Observational circumstances.}
	\label{Table1}
	\begin{tabular}{ccccc}
		\hline
		\hline
		Object&Date UT&r, AU&$\Delta$, AU&$\alpha$, deg\\
		\hline	
		(55636) 2002 TX300&2009 Aug 25.05&41.518	&40.898&1.11\\
		&2009 Aug 26.09&41.518&40.886&1.10\\
		(120178) 2003 OP32&2009 Aug 22.02&41.454&40.484&0.40\\
		&2009 Aug 24.00&41.455&40.486&0.40\\
		(145452) 2005 RN43&2009 Aug 21.92&40.695&39.706&0.30\\
		&2009 Aug 22.96&40.695&39.705&0.29\\
		&2009 Aug 24.00&40.695&39.705&0.29\\
		(148975) 2001 XA255&2009 Mar 29.06&9.566&8.616&1.93\\
		(202421) 2005 UQ513&2009 Aug 22.04&48.741&48.104&0.93\\
		&2009 Aug 23.06&48.740&48.093&0.92\\
		&2009 Aug 24.06&48.740&48.081&0.91\\
		(281371) 2008 FC76&2009 Aug  25.03&11.270&10.381&2.55\\
		&2009 Aug 25.98&11.270&10.381&2.55\\
		(315898) 2008 QD4&2009 Mar 26.10&5.878&6.842&2.34\\
		(444030) 2004 NT33&2009 Aug 24.00&38.172&37.325&0.84\\
		&2009 Aug 25.04&38.173&37.329&0.84\\
		2008 CT190&2009 Mar 28.96&34.726&34.146	&1.35\\
		\hline
	\end{tabular}
\end{table*}

\section{Results}

\begin{figure*}
	\includegraphics[width=\textwidth]{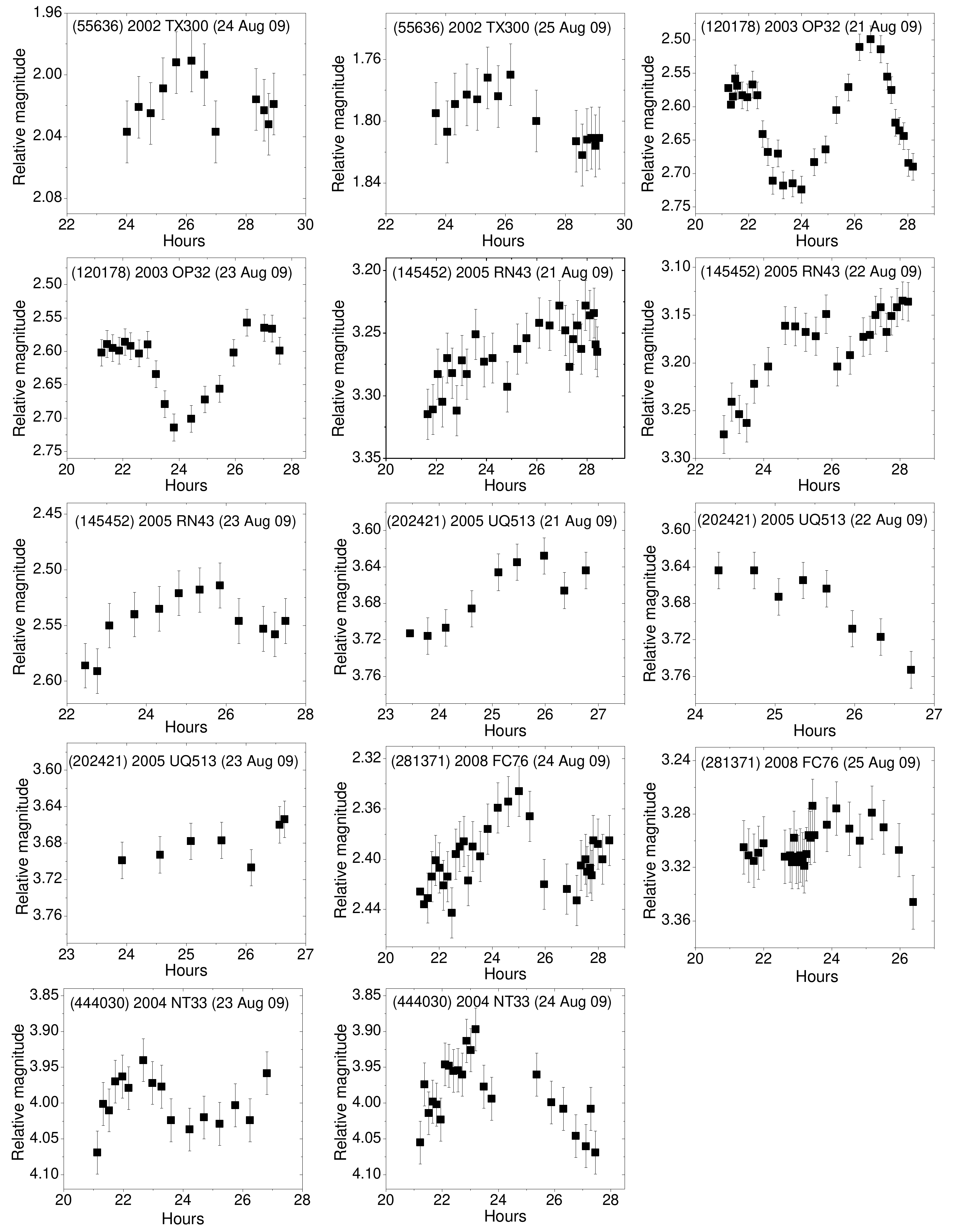}
	\caption{Individual ligthcurves of the objects that were observed for more than one night.}
	\label{ALL}
\end{figure*}

We observed 9 objects, including 5 classical TNOs (2 of them are members of the Haumea family), 
3 Centaurs and one Scattered-disk object (SDO), according to dynamical classification 
by \citet{Gladman2008}. Short-term variabilities of SDO 2008 CT190, 
Centaurs (148975) 2001 XA255, (281371) 2008 FC76, 
and (315898) 2008 QD4 were observed for the first time. The rotational periods of
 the objects that were observed for more that one night were calculated based on 
Fourier analysis technique \citep[cf.][]{Harris1989, Magnusson1990}. Figure~\ref{ALL} 
presents single-night observations, except for 2001 XA255, 2008 QD4, and 2008 CT190 
lightcurves that are shown separately.

Since each object was observed during only one opposition in order to improve the 
accuracy of rotational period values we also used literature data 
\citep[from][]{Benecchi2013, Thirouin2010, Thirouin2012} in the analysis.

We summarize in Table~\ref{Table2} previously published data on rotation periods and 
lightcurve amplitudes and our new determinations. We also give in Table~\ref{Table2} 
the orbital type of these objects, and the estimations of diameters and albedos with the 
corresponding references.  

Lower limits of densities were also derived for six objects. We used the tables
from \citet{Chandrasekhar1987} for rotationally stable Jacobi ellipsoids, and
considered the lower limits of the axial ratio \textit{a/b}. For simplicity and
given the icy-rich nature of TNOs, a fluid body (i.e. a body with no tensile
and pressure-dependent strength) assumption is normally used when calculating
the densities. We note however, that for 2008 FC76 and 2003 OP32 this approach
may not be correct, as their sizes could be too small to acquire the
hydrostatic equilibrium. For more details on the calculation of the obtained
density limit values we refer the reader to the paper \citet{Perna2009}. In
Table~\ref{Density} we provide the lower limits of the axis ratio and
estimations of the densities (assuming an elongated shape of the objects and thus the longer period with double extrema lightcurve).

A caveat is in order at this point. The assumption of
hydrostatic equilibrium (inherent in our modelling) is
plausible but clearly not ``optimal'' for the bodies under consideration. 
Objects in the TNO-Centaur population tend to be
either small enough that non-hydrostatic deviations can explain the
lightcurve amplitude (as mentioned above), or large enough that albedo 
variegation  (e.g., Pluto) can provide an explanation for the lightcurve  amplitude.

As can be seen from Table~\ref{Density} the density lower limits are extremely
low and therefore are not giving very significant information, apart 
from the, non negligible, fact that none of the observed spins 
and  lightcurve  amplitudes suggest an unexpected density.

\begin{table*}
	\centering
	\caption{Summary on the observed objects.}
	\label{Table2}
	\begin{tabular}{p{1.5cm}p{1cm}p{1cm}ccccp{1.3cm}c}
		\hline
		\hline
		\let\newline\\
		Object&Orbital type&H$^{a}, $\,mag&D, km&pv&P single, h&P double, h&A, mag&Reference\\
		\hline	
		(55636) 2002~TX300&Cl&3.3&286$^{1}$&0.88$^{1}$&8.12$\pm$0.08&16.24$\pm$0.08&0.02$\pm$0.02&\citet{Sheppard2003}\\
		&&&&&12.10$\pm$0.08&24.20$\pm$0.08&0.08$\pm$0.02&\citet{Sheppard2003}\\
		&&&&&-&15.78$\pm$0.05&0.09$\pm$0.02&\citet{Ortiz2004}\\
		&&&&&8.16$\pm$0.05&-&0.04$\pm$0.01&\citet{Thirouin2010}\\
		&&&&&8.15$\pm$0.05&-&0.05$\pm$0.01&\citet{Thirouin2012}\\
		&\\
		&&&&&8.04$\pm$0.04&16.08$\pm$0.04&0.05$\pm$0.01&This paper\\
		\hline
		(120178) 2003~OP32&Cl&4.1&$\sim$216$^{1}$&0.88$^{1}$&4.854$\pm$0.003&-&0.20$\pm$0.04&\citet{Rabinowitz2008}\\
		&&&&&4.05$\pm$0.05&-&0.13$\pm$0.01&\citet{Thirouin2010}\\
		&&&&&4.85&9.71&0.18$\pm$0.01&\citet{Benecchi2013}\\
		&&&&&4.85&&0.14$\pm$0.02&\citet{Thirouin2016}\\
		\\
		&&&&&-&9.7057$\pm$0.0001&0.15$\pm$0.01; 0.18$\pm$0.01&This paper\\	
		\hline	
		(145452) 2005~RN43&Cl&3.9&679$^{2}$&0.11$^{2}$&5.62$\pm$0.05&-&0.04$\pm$0.01&\citet{Thirouin2010}\\
		&&&&&6.95$\pm$0.05&13.89$\pm$0.05&0.06$\pm$0.01&\citet{Benecchi2013}\\
		&\\
		&&&&&6.946$\pm$0.05&13.892$\pm$0.05&0.04$\pm$0.01&This paper\\
		\hline
		(148975) 2001~XA255&Cen&11.1&38$^{3}$&0.04$^{3}$&> 7&> 14&$\sim$0.2&This paper\\
		\hline
		(202421) 2005~UQ513&Cl&3.4&498$^{4}$&0.20$^{4}$&7.03$\pm$0.05&-&0.06$\pm$0.02&\citet{Thirouin2012}\\
		&\\
		&&&&&7.03$\pm$0.005&14.06$\pm$0.005&0.07$\pm$0.01&This paper\\
		\hline
		(281371) 2008~FC76&Cen&9.3&$\sim$41-82$^{5}$&-&5.93$\pm$0.05&11.86$\pm$0.05&0.04$\pm$0.01&This paper\\
		\hline
		(315898) 2008~QD4&Cen&11.3&$\sim$16-35$^{5}$&-&> 7&> 14&$\sim$0.15&This paper\\
		\hline
		(444030) 2004~NT33&Cl&4.7&423$^{4}$&0.13$^{4}$&57.87.$\pm$0.05&-&0.04$\pm$0.01&\citet{Thirouin2012}\\
		&\\
		&&&&&7.871$\pm$0.05&15.742$\pm$0.05&0.05$\pm$0.01&This paper\\
		\hline
		2008~CT190&SDO&5.5&$\sim$236-470$^{5}$&-&> 5&> 10&$\sim$0.15&This paper\\
		\hline
	\end{tabular}
	\begin{flushleft}
	$^{a}$From Minor Planet Centre  database, \\
	$^{1}$\citet{Elliot2010},
	$^{2}$\citet{Vilenius2012},
	$^{3}$\citet{Braga-Ribas2012},
	$^{4}$\citet{Vilenius2014},
	$^{5}$Assuming an albedo range of 0.05-0.20.
	\end{flushleft}
\end{table*}

\begin{table}
	\centering
	\caption{Lower limits of axis ratio and density values (together with diameter and periods, from 
Table~\ref{Table2}). See text for details.}
	\label{Density}
	\begin{tabular}{ccccc}
		\hline
		\hline
		\let\newline\\
		Object&axis ratio &$\rho$ & Diameter&Period\\
                      & \textit{a/b} & [g cm$^{-3}$] & [km] & single/double [h] \\
		\hline

		(55636) &1.05&0.15&286&8.04/16.08\\
		(120178)&1.18&0.41&$\sim$ 108&-/9.706\\
		(145452)&1.04&0.20&679&6.95/13.892\\
		(202421)&1.07&0.20&498&7.03/14.06\\
		(281371)&1.04&0.28&$\sim$41-82&5.93/11.86\\
		(444030)&1.05&0.16&423&7.871/15.742\\
		\hline
	\end{tabular}
\end{table}

\subsection{(55636) 2002 TX300}

(55636) 2002 TX300 is a classical TNO, which is one of the largest member of the Haumea family.
 The object has highly inclined orbit similar to that of (136108) Haumea. Previously reported values
 of its rotational period vary between 8.12 and 24.2~h (considering both single and double-peaked lightcurves) 
with estimated amplitude of about 0.04-0.09~mag  \citep{Sheppard2003, Ortiz2004, Thirouin2010, Thirouin2012}.

\begin{figure}
	\includegraphics[width=\columnwidth]{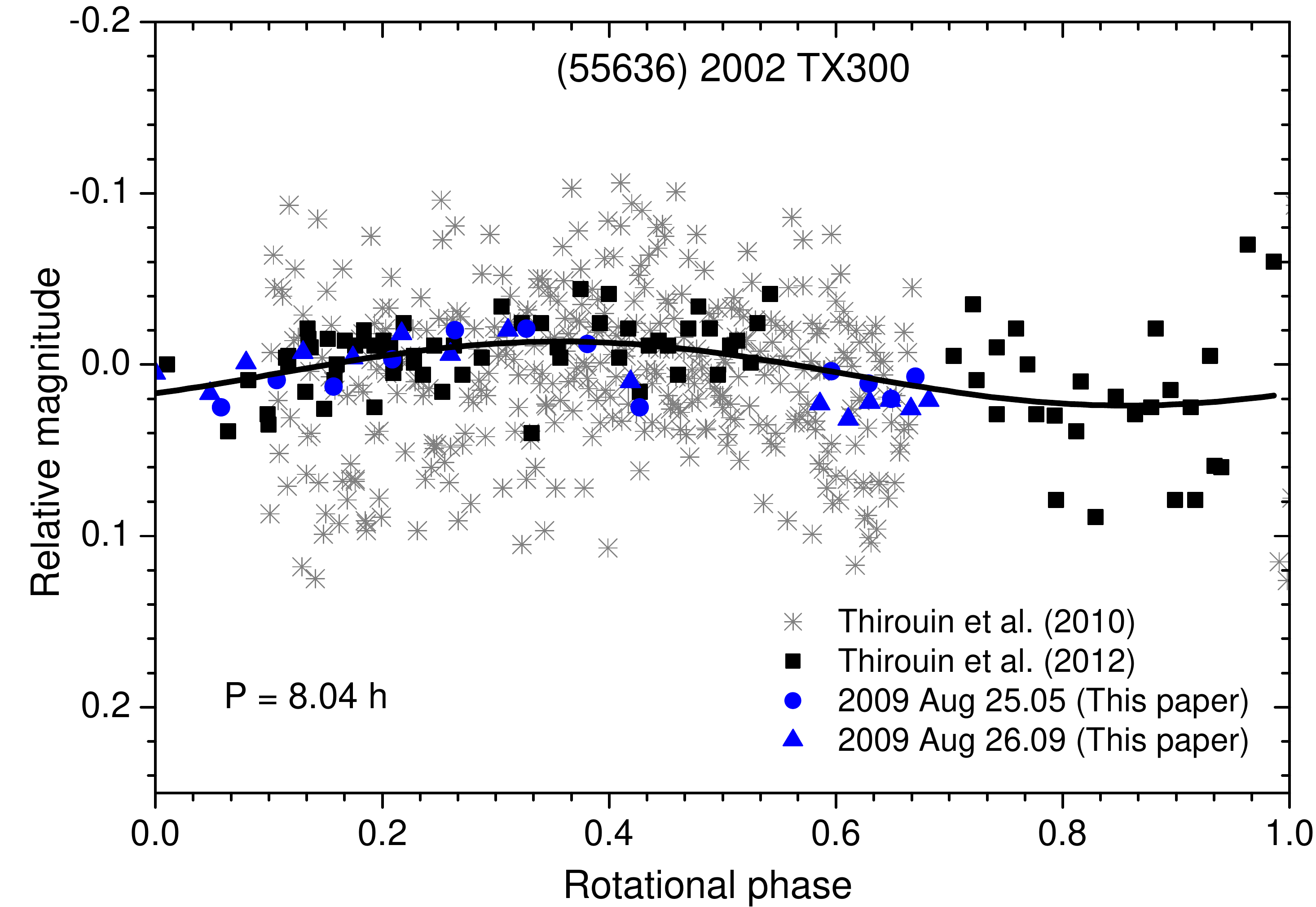}
    \caption{Single-peaked composite lightcurve of (55636) 2002 TX300. Zero phase is at UT August 25.7497, 2009.}
    \label{lc2002tx300_short_all}
\end{figure}

\begin{figure}
	\includegraphics[width=\columnwidth]{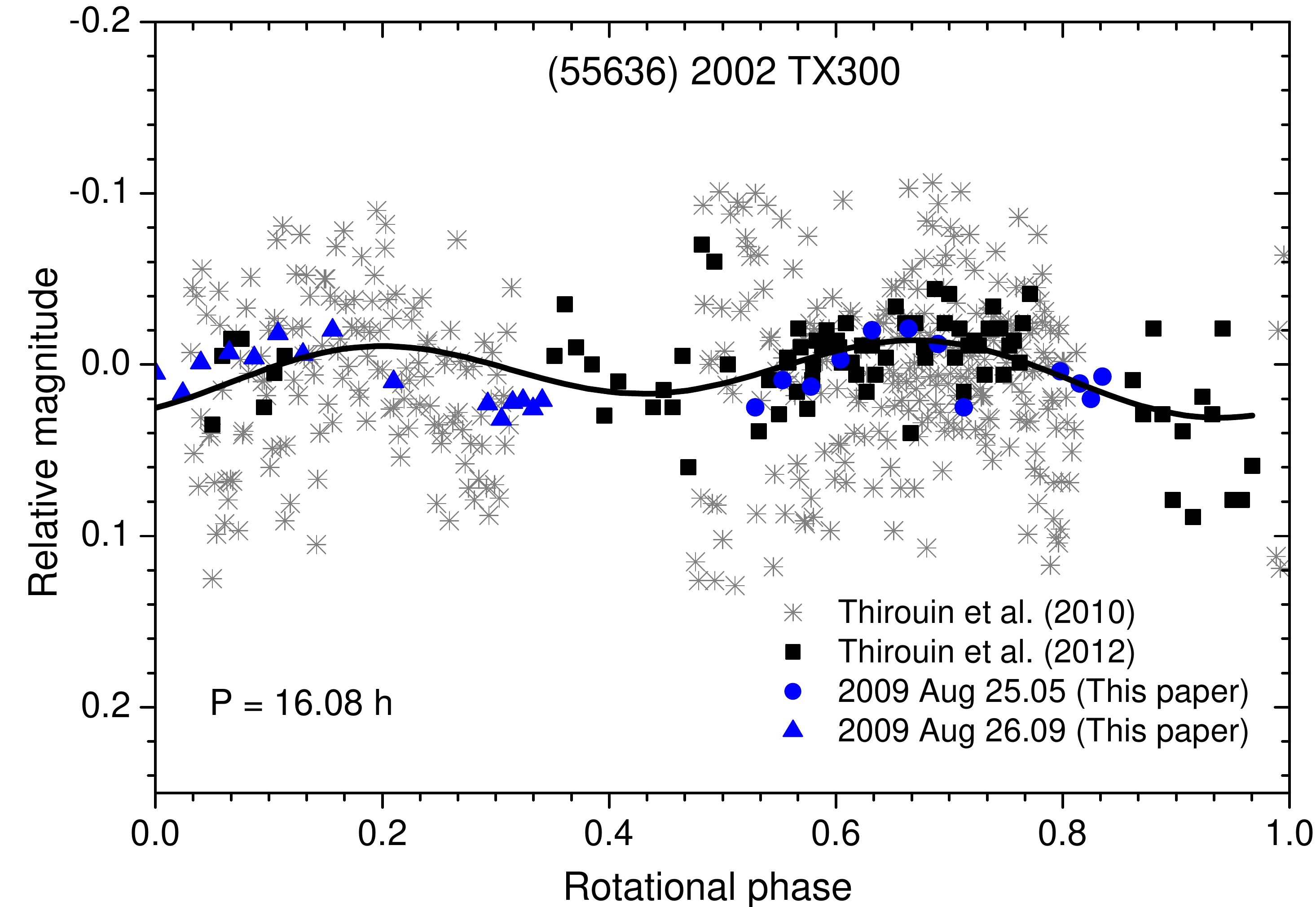}
    \caption{Double-peaked composite lightcurve of (55636) 2002 TX300. Zero phase is at UT August 25.7497, 2009.}
    \label{lc2002tx300_long_all}
\end{figure}

(55636) 2002 TX300 was observed for two consequent nights on August  24-25, 2009. 
We found the rotational periods 8.04$\pm$0.04~h and its double value 16.08$\pm$0.04~h 
with the amplitude of 0.05$\pm$0.01~mag. The composite lightcurves for single and double-peaked
 rotational periods are shown in Fig.~\ref{lc2002tx300_short_all} and  Fig.~\ref{lc2002tx300_long_all} respectively.

\subsection{(120178) 2003 OP32}
\label{(120178) 2003 OP32}

(120178) 2003 OP32 is another member of the Haumea collisional family that we observed. It also has orbital parameters similar to that of Haumea. This object was previously observed by different authors that report rotational periods from 4.05 to 9.71~h \citep{Rabinowitz2008, Thirouin2010, Benecchi2013, Thirouin2016} and amplitude from 0.13 to 0.20~mag. The authors of above mentioned papers did not report a lightcurve asymmetry, and could not give a preference to single or double-peaked variability rate.

\begin{figure}
	\includegraphics[width=\columnwidth]{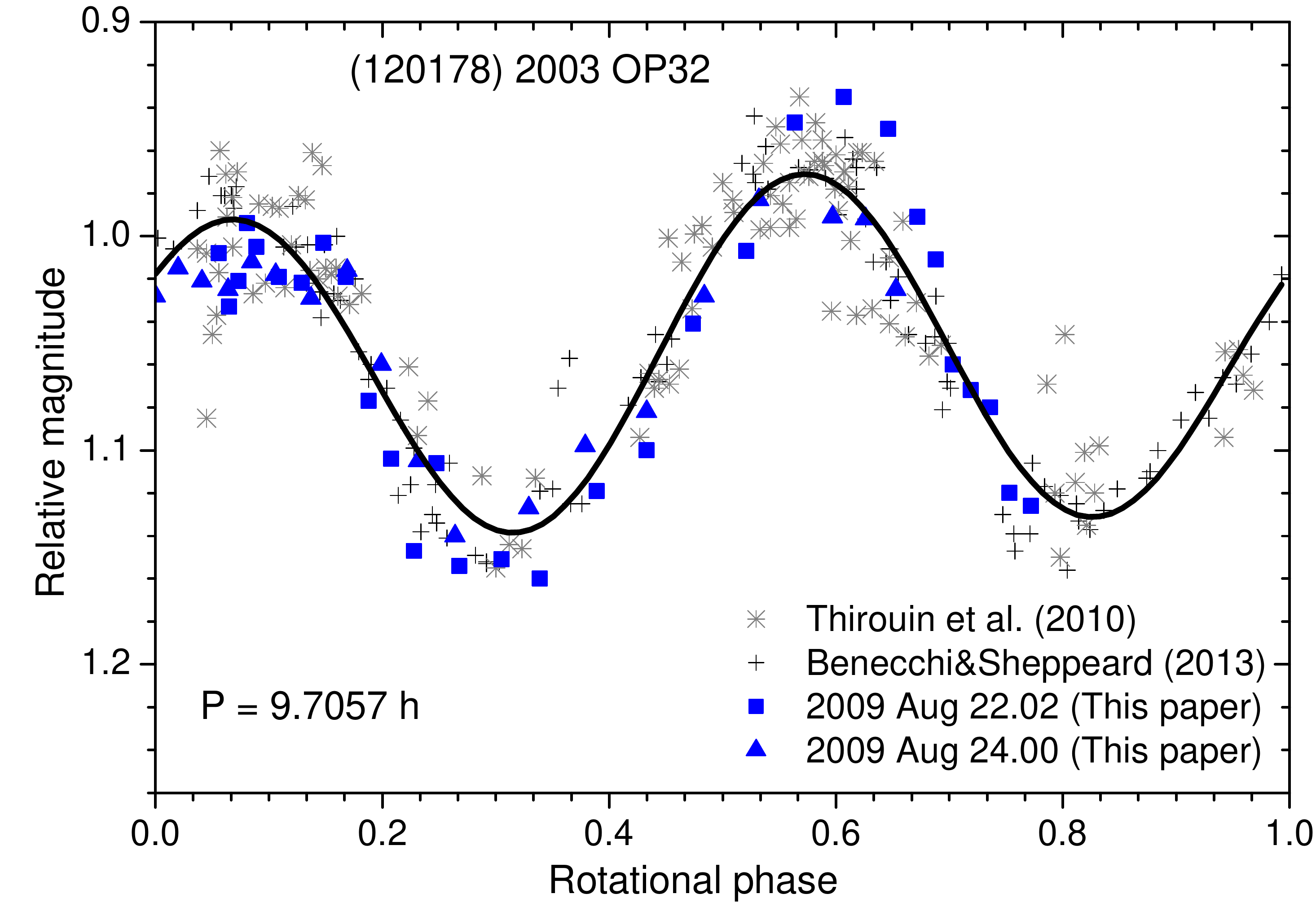}
    \caption{Double-peaked composite lightcurve of (120178) 2003 OP32. Zero phase is at UT August 23.6512, 2009.}
    \label{lc2003op32_all}
\end{figure}

We observed this object for two nights on August 21, 23, 2009. Our results show a difference in amplitude of about 0.03~mag between two peaks, suggesting double-peaked period. We used previously published data to both check our assumption on asymmetry and improve the precision of the rotational period. The composite lightcurve using all available to us literature data is shown in Fig.~\ref{lc2003op32_all}.
The rotation period is 9.7057$\pm$0.0001~h with the primary amplitude 0.18$\pm$0.01~mag and secondary 0.15$\pm$0.01~mag.

\subsection{(145452) 2005 RN43}

(145452) 2005 RN43 is a classical Kuiper belt object on a moderately inclined and almost circular orbit. It was observed previously by \citet{Thirouin2010} and \citet{Benecchi2013}. They found rotational periods of 5.62~h and 6.95~ h respectively with quite small lightcurve amplitude of about 0.05~mag.

\begin{figure}
	\includegraphics[width=\columnwidth]{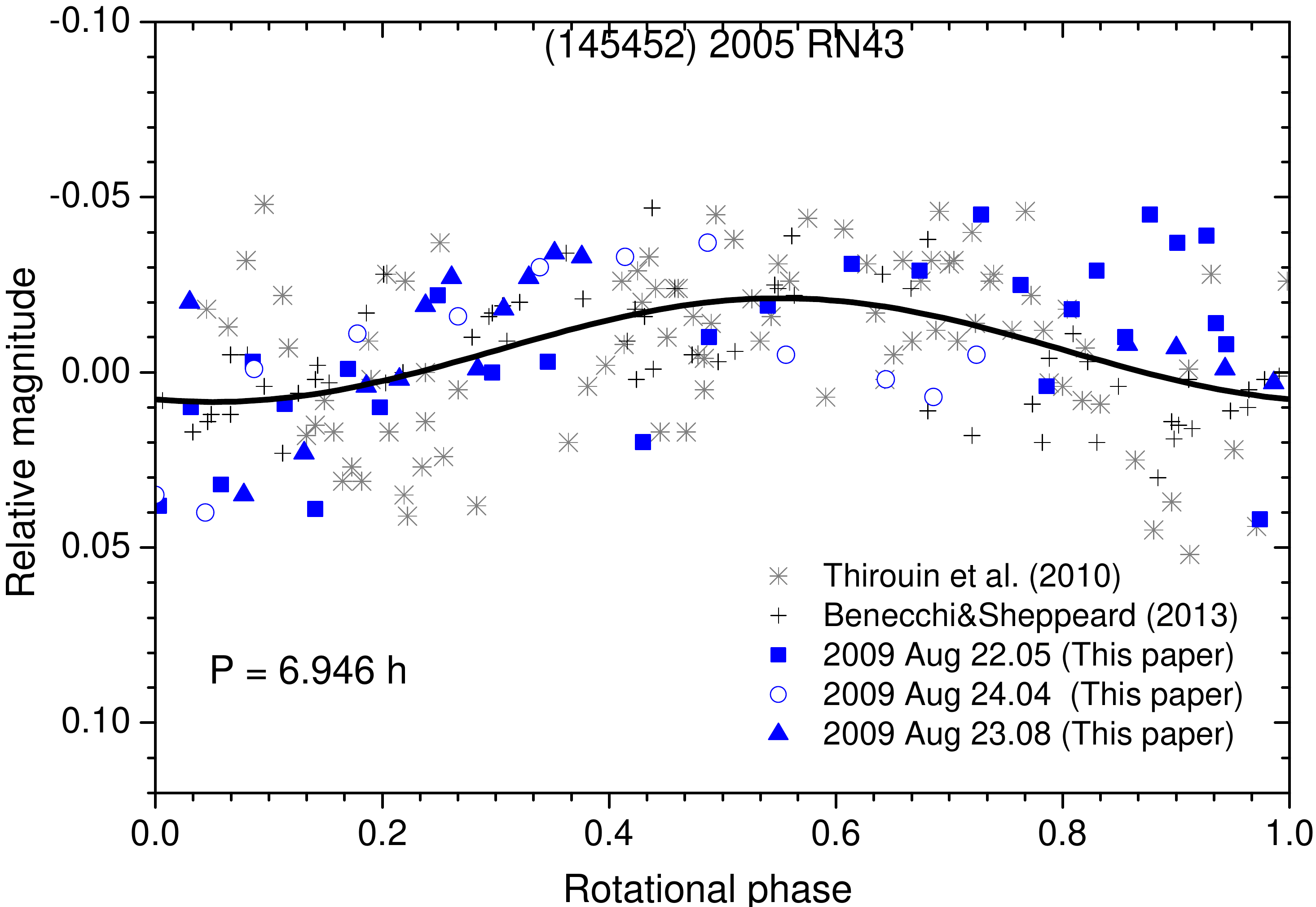}
    \caption{Single-peaked composite lightcurve of (145452) 2005 RN43. Zero phase is at UT August 23.7067, 2009.}
    \label{lc2005rn43_short_all}
\end{figure}

\begin{figure}
	\includegraphics[width=\columnwidth]{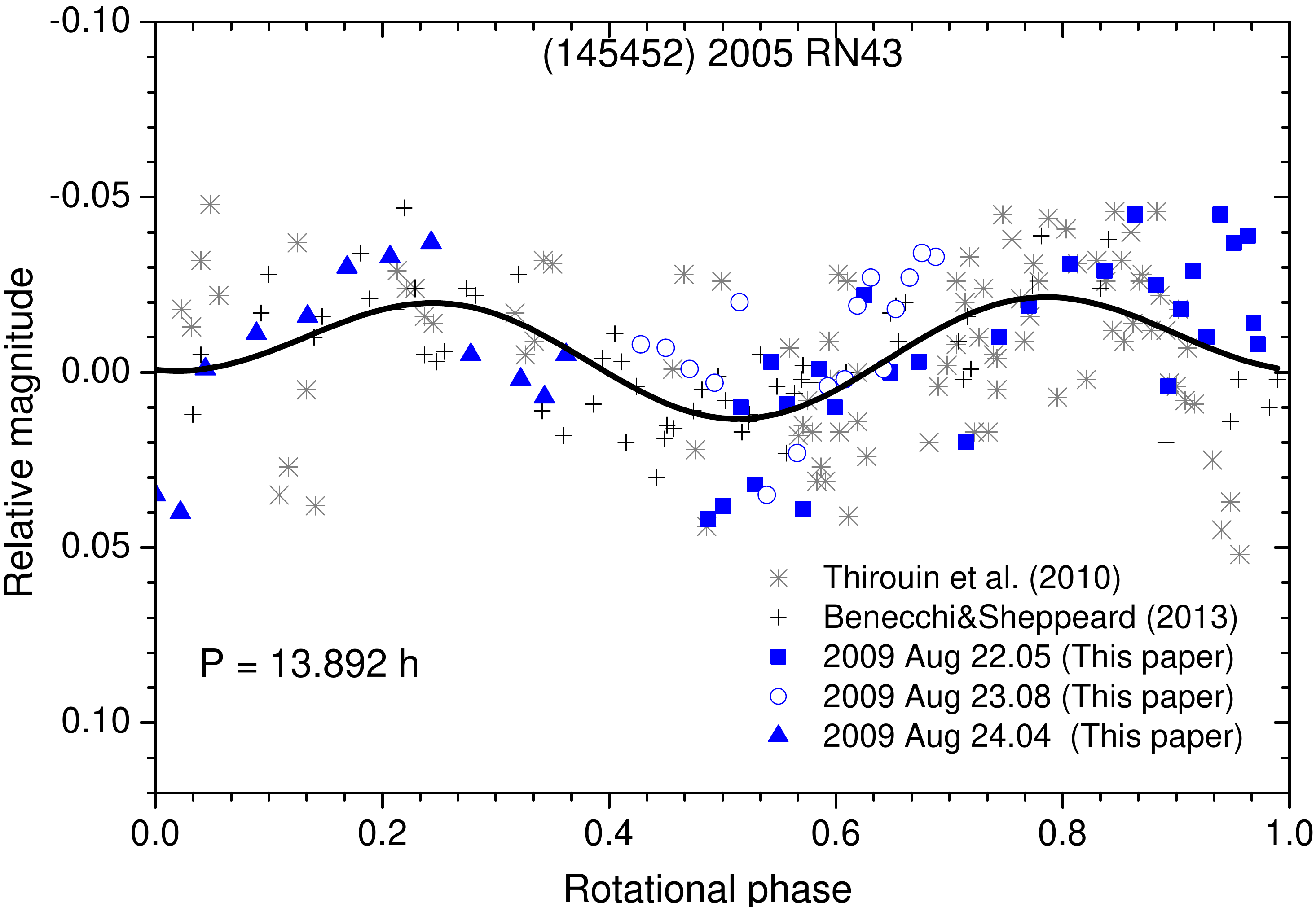}
    \caption{Double-peaked composite lightcurve of (145452) 2005 RN43. Zero phase is at UT August 23.7067, 2009.}
    \label{lc2005rn43_long_all}
\end{figure}

The observations of this TNO were performed during three consequent nights on August 21-23, 2009. From our and published data we found rotational periods of 6.946$\pm$0.05~h for a single-peaked lightcurve (Fig.~\ref{lc2005rn43_short_all}) and 13.892$\pm$0.05~h for a double-peaked lightcurve  (Fig.~\ref{lc2005rn43_long_all}). The lightcurve amplitude is 0.04$\pm$0.01~mag.

\subsection{(148975) 2001 XA255}

\begin{figure}
	\includegraphics[width=\columnwidth]{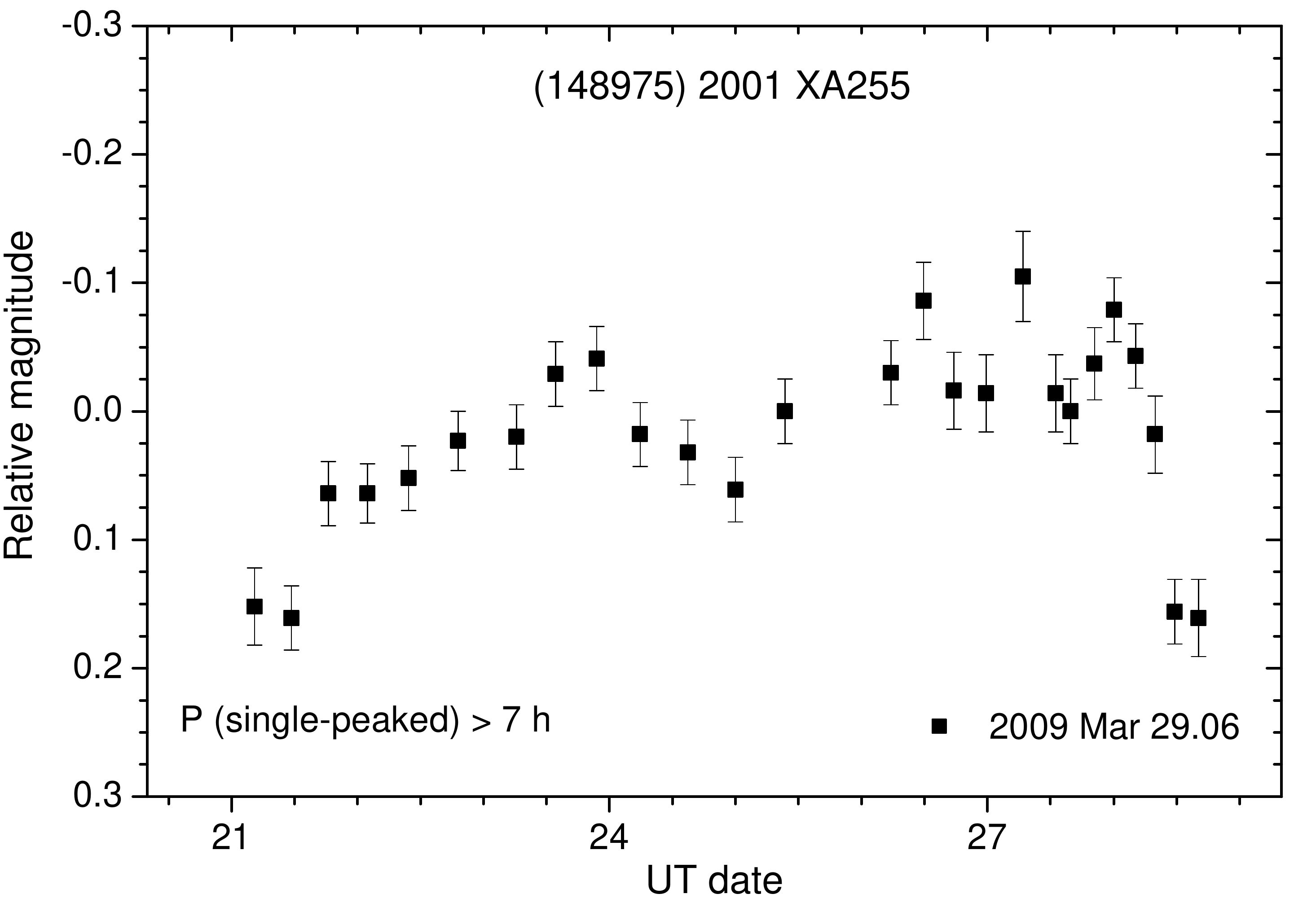}
    \caption{Lightcurve of (148975) 2001 XA255.}
    \label{lc2001xa255}
\end{figure}

(148975) 2001 XA255 was first classified as a Centaur, but later \citet{Fuente2012} suggested that this object is a dynamically unstable temporary Neptune co-orbital. Authors argue it may be a relatively recent visitor from the scattered disk on its way to the inner Solar system. No rotational period values are reported in the literature. As we observed 2001 XA255 during only one night on March 28, 2009, we can just give a lower limit of a rotational period to be about 7~h (or 14~h if double-peaked) with an amplitude $\sim$0.2~mag (Fig.~\ref{lc2001xa255}).

\subsection{(202421) 2005 UQ513}

(202421) 2005 UQ513 is a classical TNO. \citet{Thirouin2012} reported a rotational variability of 7.03~h and quite small amplitude of 0.06~mag. 

\begin{figure}
	\includegraphics[width=\columnwidth]{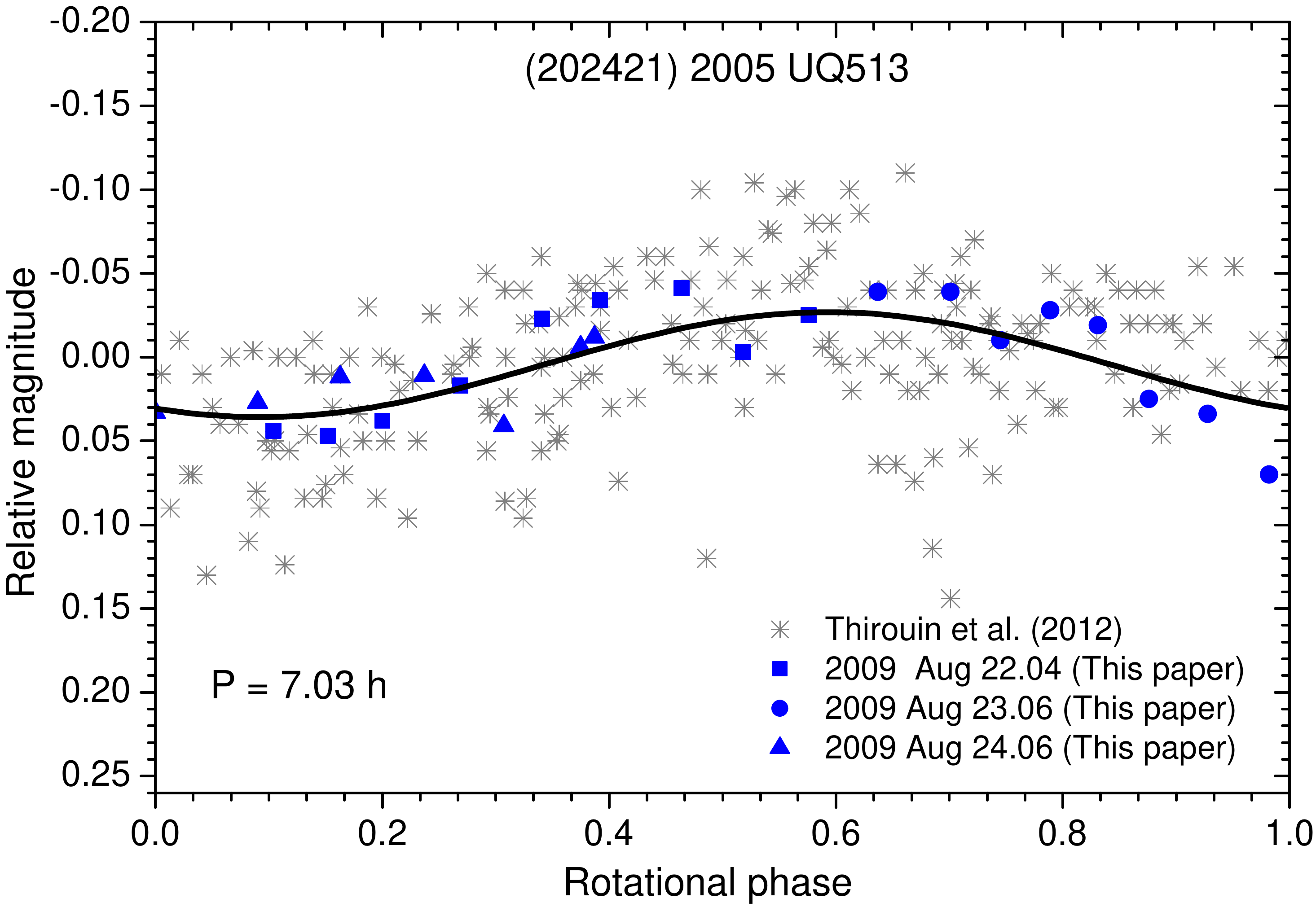}
    \caption{Single-peaked composite lightcurve of (202421) 2005 UQ513. Zero phase is at UT August 23.7193, 2009.}
    \label{lc2005uq513_short_all}
\end{figure}

\begin{figure}
	\includegraphics[width=\columnwidth]{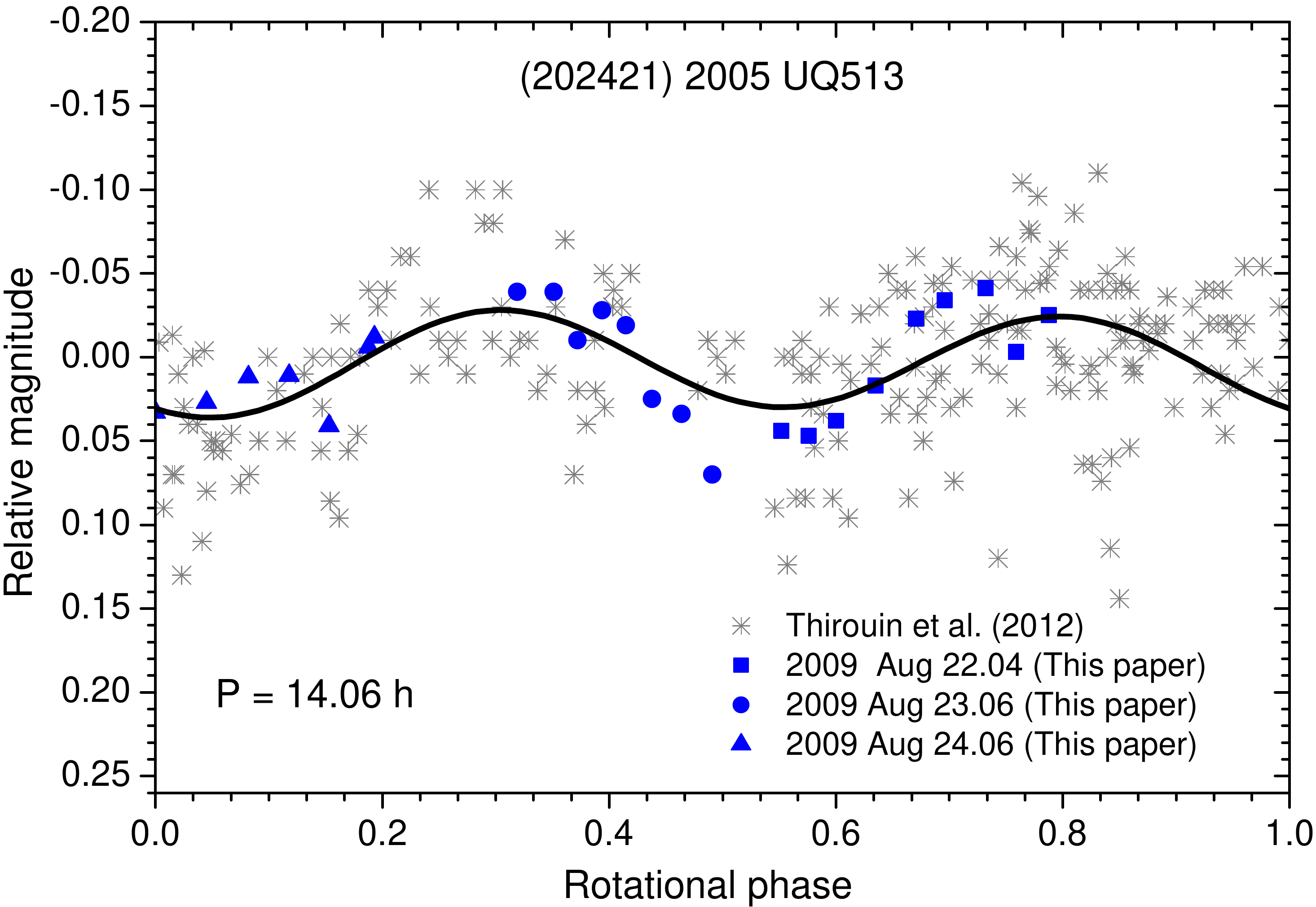}
    \caption{Double-peaked composite lightcurve of (202421) 2005 UQ513. Zero phase is at UT August 23.7193, 2009.}
    \label{lc2005uq513_long_all}
\end{figure}

From our observations during three nights on August 21-23, 2009 we can confirm this value and suggest 7.03$\pm$0.05~h (single-peaked) and 14.06$\pm$0.05~h (double-peaked) short-term variability with an amplitude 0.07$\pm$0.01~mag. The single and double-peaked composite lightcurves for this object are presented in Fig.~\ref{lc2005uq513_short_all} and  Fig.~\ref{lc2005uq513_long_all} respectively.

\subsection{(281371) 2008 FC76}

(281371) 2008 FC76 is a Centaur on a moderately eccentric and highly inclined orbit. No short-term variability values are available in the literature.

\begin{figure}
	\includegraphics[width=\columnwidth]{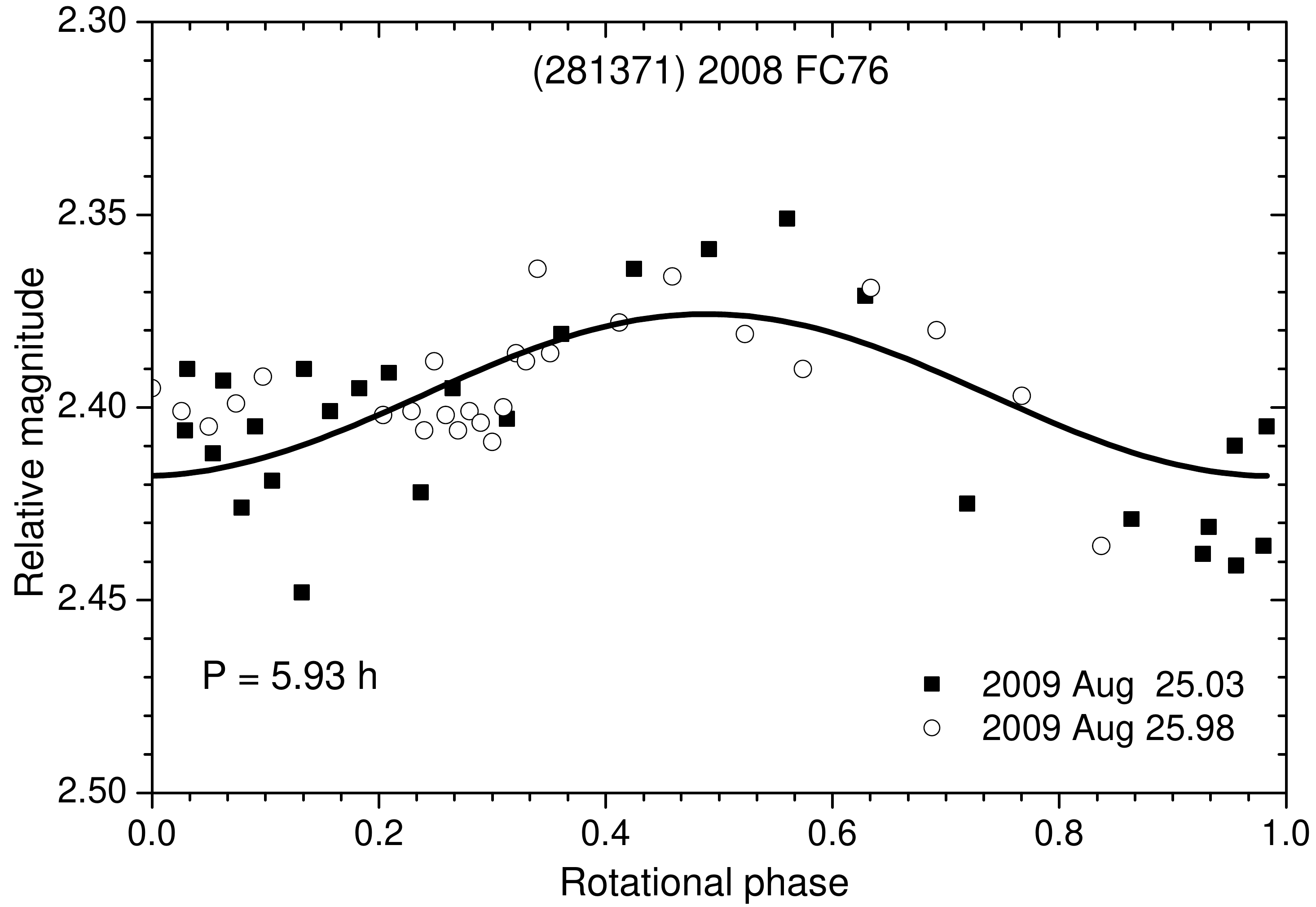}
    \caption{Single-peaked composite lightcurve of (281371) 2008 FC76. Zero phase is at UT August 25.8323, 2009.}
    \label{lc2008fc76_short}
\end{figure}

\begin{figure}
	\includegraphics[width=\columnwidth]{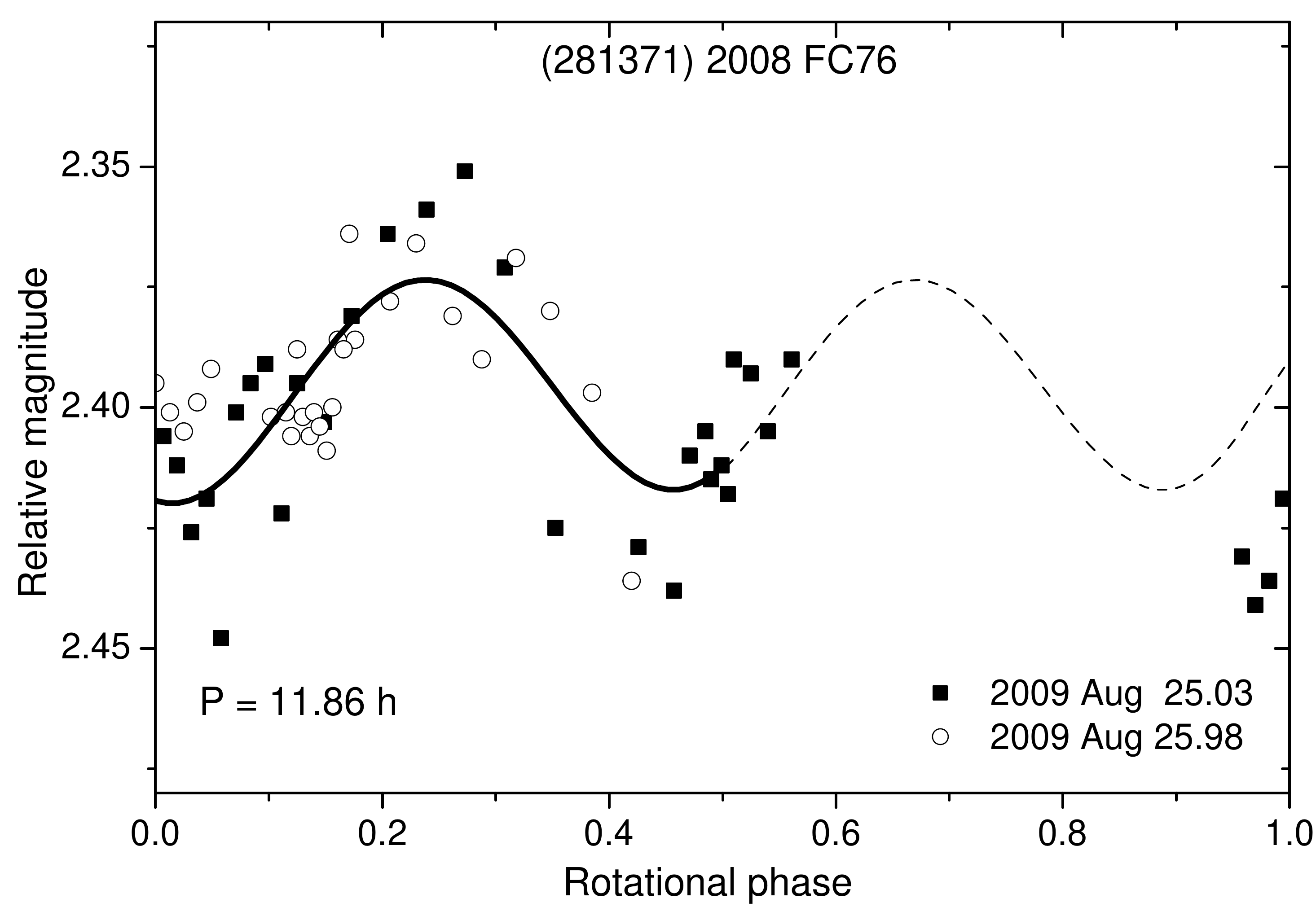}
    \caption{Double-peaked composite lightcurve of (281371) 2008 FC76. Zero phase is at UT August 25.8323, 2009.}
    \label{lc2008fc76_long}ysxll
\end{figure}

We observed this object during two nights on August 24-25, 2009. Our data suggest a rotational period of 5.93$\pm$0.05~h (or twice this value 11.86$\pm$0.05~h) with peak-to-peak variation of 0.04$\pm$0.01~mag. Composite lightcurves for single and double-peaked solutions are shown in Fig.~\ref{lc2008fc76_short} and Fig.~\ref{lc2008fc76_long} respectively.

\subsection{(315898) 2008 QD4}

\begin{figure}
	\includegraphics[width=\columnwidth]{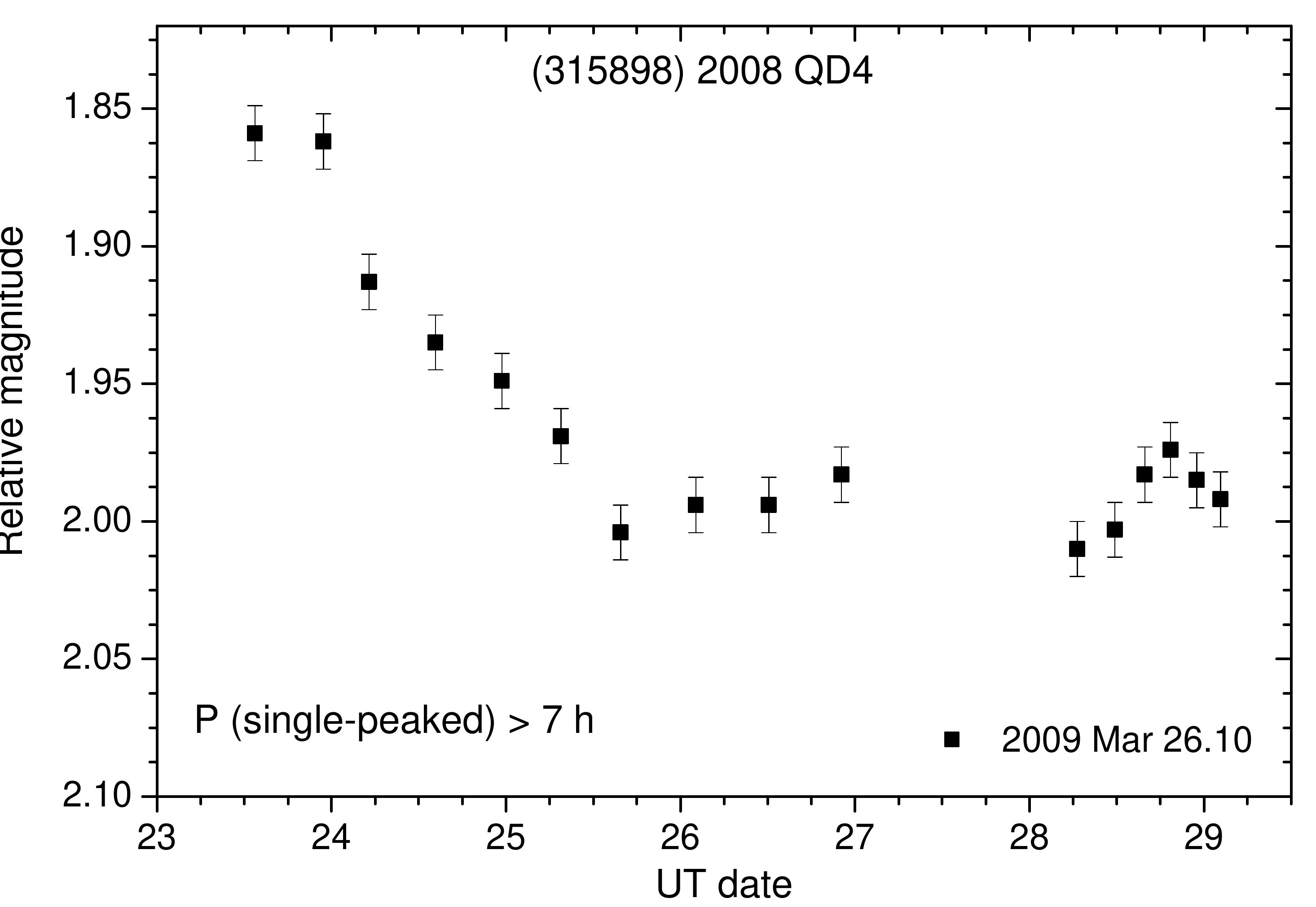}
    \caption{Lightcurve of (315898) 2008 QD4.}
    \label{lc2008qd4}
\end{figure}

(315898) 2008 QD4 has been classified as a Centaur on a highly eccentric and highly inclined orbit. No values of its rotational period were reported so far. Based on one-night observations on March 25, 2009 (Fig.~\ref{lc2008qd4}) we suggest its rotational period to be longer than $\sim$7~h and lightcurve amplitude value of about $\sim$ 0.15~mag.

\subsection{(444030) 2004 NT33}

\begin{figure}
	\includegraphics[width=\columnwidth]{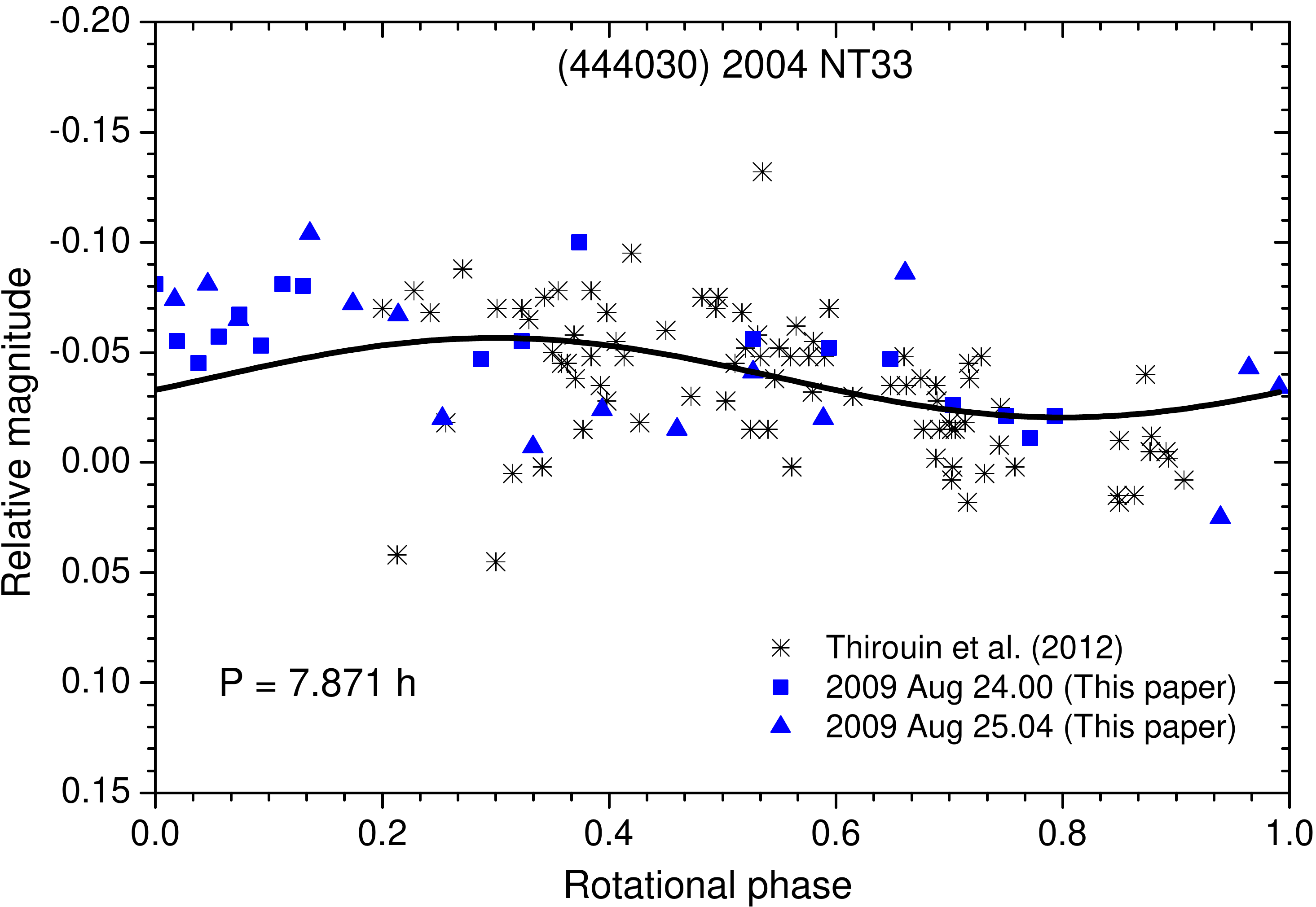}
    \caption{Single-peaked composite lightcurve of (444030) 2004 NT33. Zero phase is at UT August 24.6685, 2009.}
    \label{lc2004nt33_short_all}
\end{figure}

\begin{figure}
	\includegraphics[width=\columnwidth]{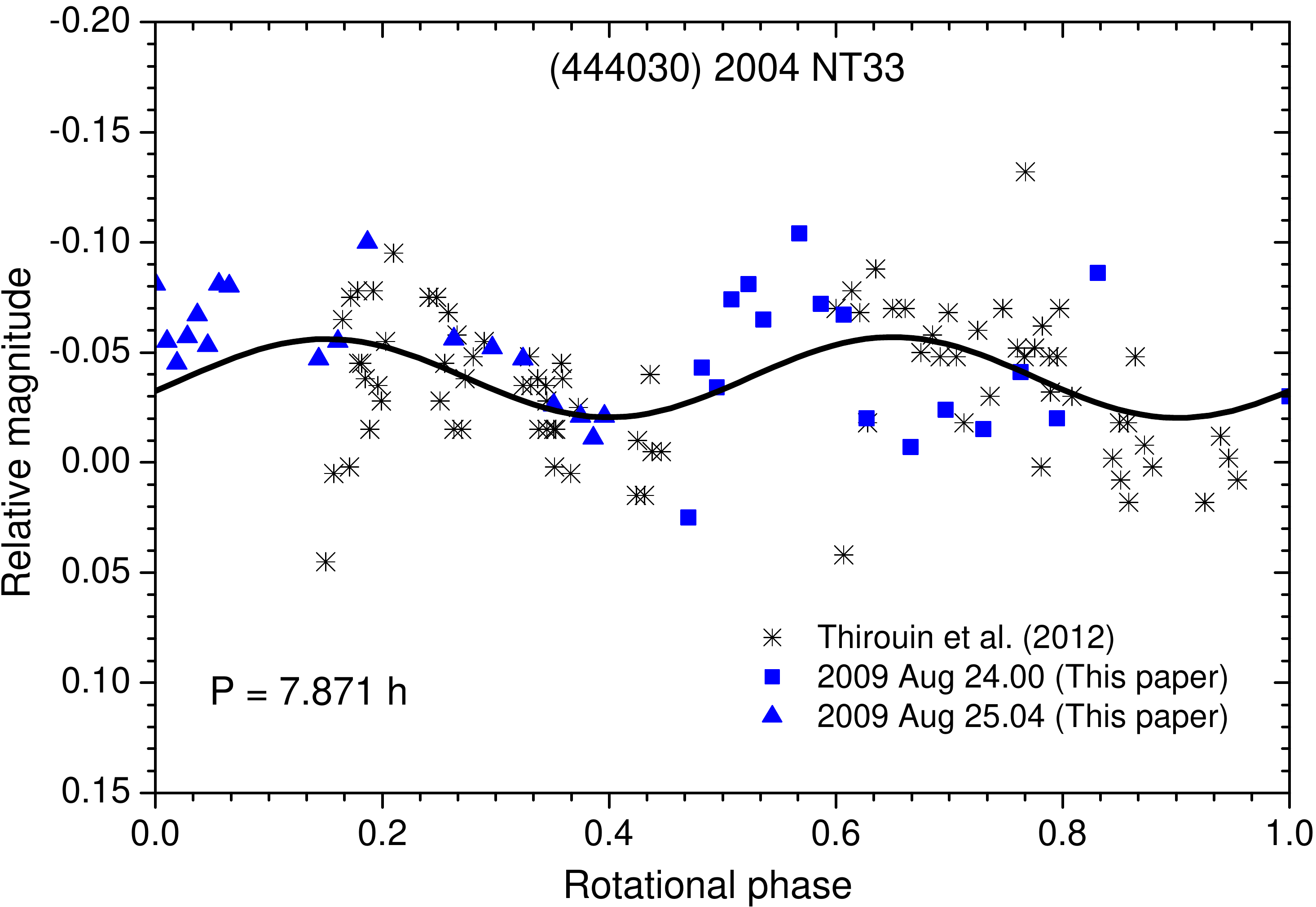}
    \caption{Double-peaked composite lightcurve of (444030) 2004 NT33. Zero phase is at UT August 24.6685, 2009.}
    \label{lc2004nt33_long_all}
\end{figure}

(444030) 2004 NT33 is a classical, low-eccentricity and high-inclination TNO. Previously reported value of its rotation period is 7.87~h (single-peaked light-curve), and suggested amplitude is quite low (0.04~mag) \citep{Thirouin2012}.
We observed this object on August 23-24, 2009. From our and literature data we found more accurate rotational period value of 7.871$\pm$0.05~h (15.742$\pm$0.05~h for double-peaked period) that is consistent with all of the available data and has an amplitude of 0.05$\pm$0.01~mag. Fig.~\ref{lc2004nt33_short_all} and  Fig.~\ref{lc2004nt33_long_all} show the composite lightcurves for single and double-peaked lightcurves respectively.

\subsection{2008 CT190}
\begin{figure}
	\includegraphics[width=\columnwidth]{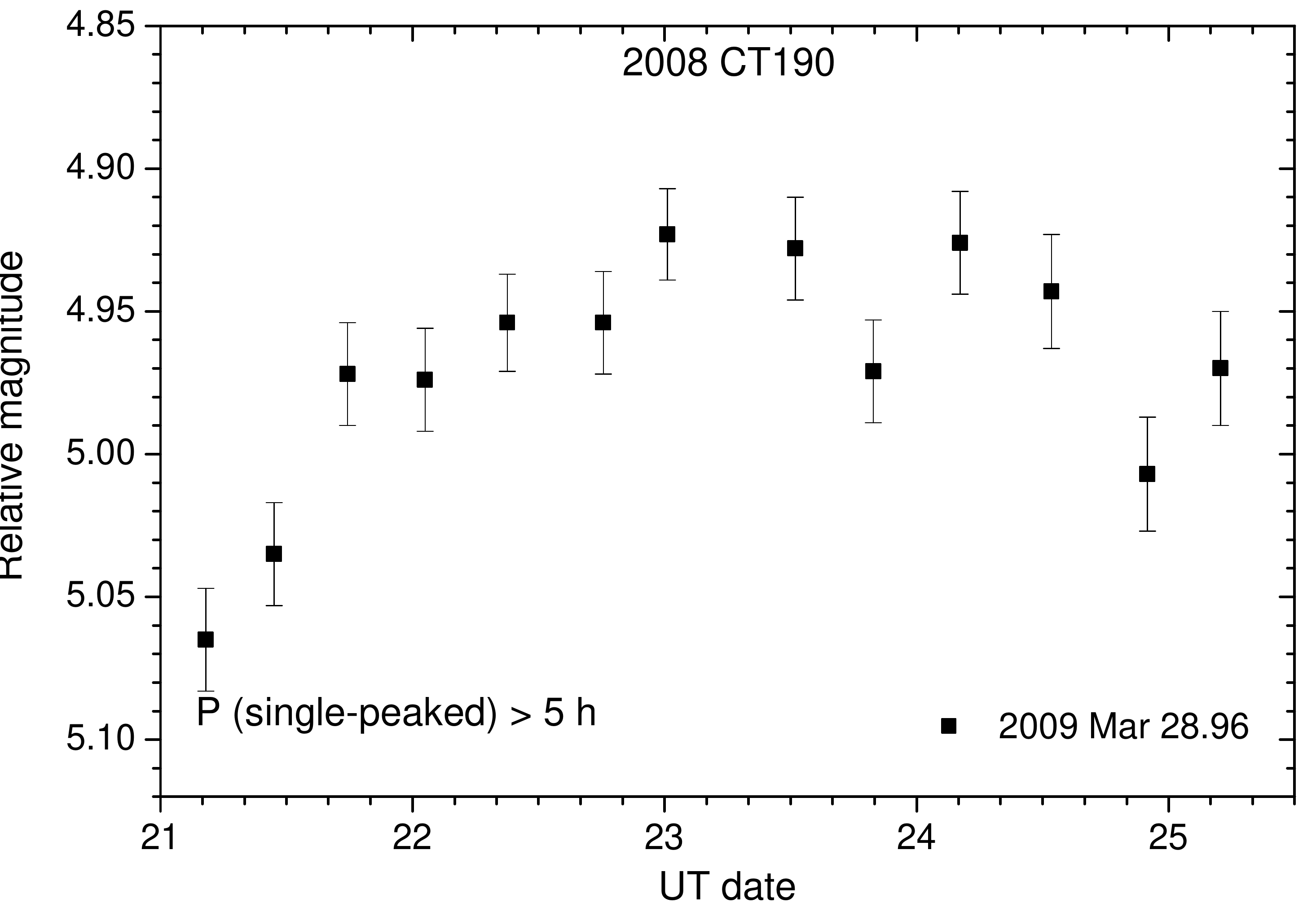}
    \caption{Lightcurve of 2008 CT190.}
    \label{lc2008ct190}
\end{figure}

2008 CT190 is a high-eccentricity and high-inclination TNO. This object is less studied compared to other objects in our sample. No rotational period was reported previously. From our data obtained only during 4.5 hours on March 28, 2009 we suggest a lower limit of rotational period value to be $\sim$5~h with an amplitude of $\sim$0.15~mag (Fig.~\ref{lc2008ct190}).

\section{Discussion}
\label{Discussion}

The main causes of short-term photometric variability of small Solar system bodies are aspherical shape and surface albedo variations. Lightcurve with one pair of extrema can be produced only by some kind of surface heterogeneity, whereas lightcurves with two pairs are usually associated with elongated shape.

It was shown that albedo variations contribution into rotational lightcurve is
relatively small (less than $\sim$0.2~mag).
For bodies with larger amplitudes we postulated an 
elongated shape  with double-peaked rotational periods
\citep[see][and references therein]{Thirouin2016}. As noted before,
the underlying assumption based on Jacobi ellipsoids, is at the
limits of its applicability in the case of large TNOs and 
Centaurs and is generally not valid for smaller objects (and
asteroids in general), where non-hydrostatic deviations in shape
are generally responsible for the lightcurve amplitudes.

Indeed, any lightcurve asymmetry can be caused by shape and/or albedo irregularities. 
For the majority of asteroid lightcurves the primary and secondary peaks differ 
in amplitude which let us to identify double-peaked periods. In the case of TNOs relatively large rotational
periods together with small amplitudes do not let to make a confident distinguish between 
two peaks. In our sample we found a confident evidence for double-peaked period for only one object.

However, in the case of symmetrical low-amplitude lightcurves it is more tricky to distinguish between single and double-peaked periods. Peak-to-peak values gradually decrease and completely disappear while object is approaching pole-on aspect (angle between the rotational axis and the line of sight is 0$^{\circ}$). Thus,  bodies with low amplitude lightcurves can be either viewed from near pole-on orientation or have almost spherical shapes (MacLaurin spheroids). Assuming random rotational axis orientation distribution \citet{Sheppard2002} showed that the average viewing angle would be 60$^{\circ}$, and near-spherical shapes of low-amplitude bodies are more probable, than polar observing aspect. 

It was shown by \citet{Sheppard2002} that a population of outer Solar system objects tend to be statistically more elongated than that in the Main belt. For objects with D>200~km about $\sim$30$\%$ and $\sim$23$\%$ of TNOs have lightcurve amplitudes of more than 0.15 and 0.40~mag respectively, compared to the $\sim$11$\%$ with amplitude more than 0.40~mag for the Main-belt asteroids \citep{Romanishin1999, Sheppard2002}. This may be caused by generally higher angular momentum in the Kuiper belt. Noteworthy, the majority of Main-belt asteroids are found to have double-peaked lightcurves caused by elongated shape \citep[e.g.][]{Marchis2006, Chiorny2007, Shevchenko2009, Szabo2016}.

Moreover, there is a certain correlation between the lightcurve amplitude and solar phase angle, i.e. lightcurve amplitudes tend to be smaller at smaller phase angles, and lightcurve shape effects are more pronounced at larger phase angles of about 20$^{\circ}$ \citep{Zappala1990, Kaasalainen2001}. And indeed, from ground-based sites TNOs can be observed only at small solar phase angles of less than a few degrees. As a result, TNOs with the same elongation would have smaller amplitude compared to that of a Main-belt asteroid. We also would like to emphasize that as Main-belt asteroids are closer to the observer than TNOs, their aspects of observations are changing a lot, and therefore it is easier to detect shape irregularities. Thus, considering these points, we suggest that more distant small bodies are also tend to have aspherical shapes.

From our data sample (which is quite limited) four out of nine (44$\%$) objects
have an amplitude larger or about 0.15~mag and have lightcurves that can only
be caused by elongated shape. For the rest of the objects we detect quite small
amplitudes. However, as it was shown in \citet{Johansen2012} objects with sizes
D<200~km in the Main belt and with D<300~km in the Kuiper belt cannot go
through process of self-gravitation and acquire spherical shape. In our sample
we have five objects that fall into that category, and two of them have
amplitudes smaller than 0.15~mag. We did not find any correlations between
rotational and orbital properties, though our data sample is quite small and
further investigations on this are needed.

\begin{figure}
	\includegraphics[width=\columnwidth]{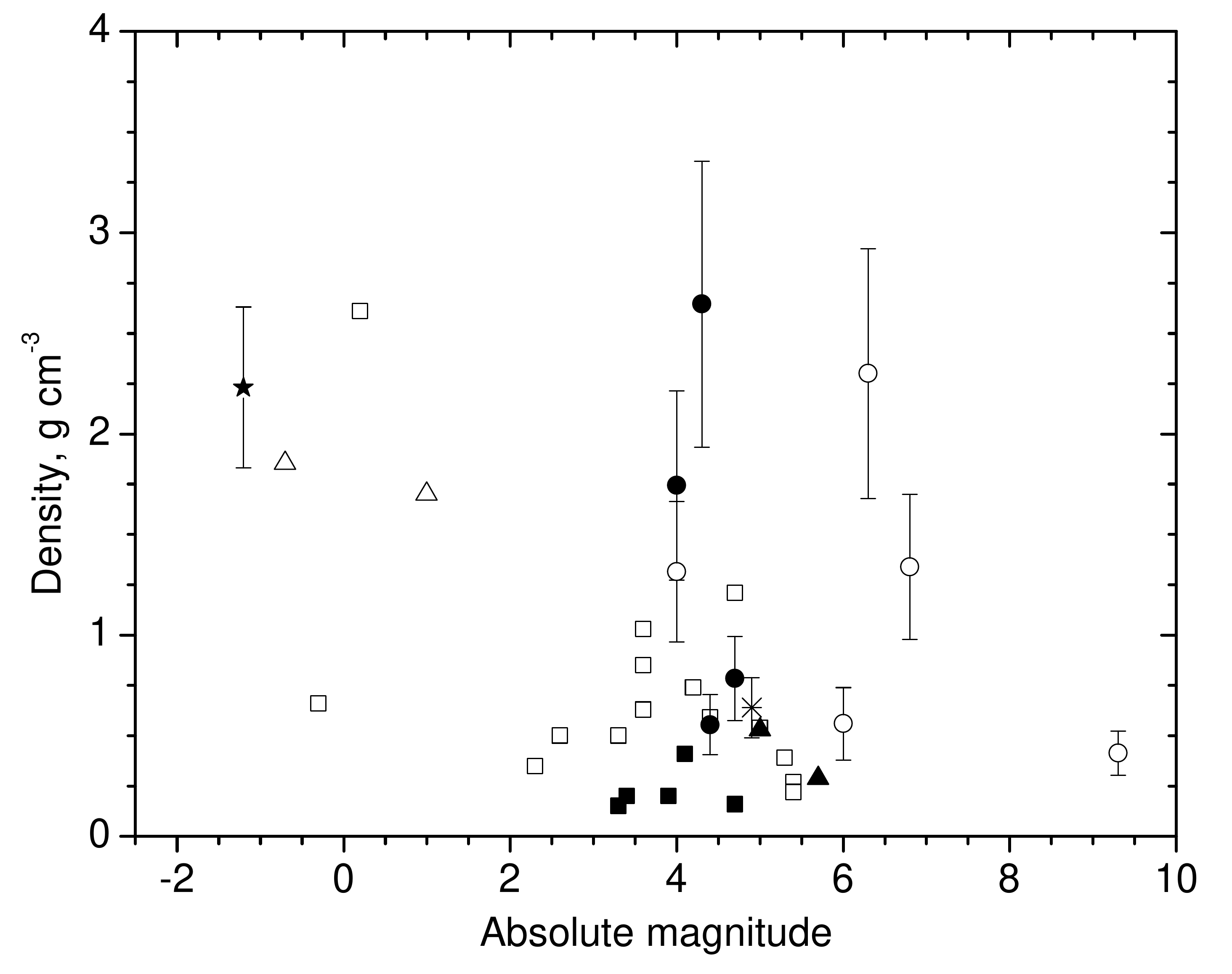}
    \caption{Density esmations of TNOs and Centaurs as a function of their absolute
magnitude. Data from this work (filled squares), literature data from \citet{Dotto2008}, open circles; \citet{Perna2009}, filled circles; \citet{Thirouin2010}, open squares; \citet{Mommert2012}, asterisks; \citet{Santos-Sanz2012}, stars; \citet{Thirouin2012}, filled triangles; \citet{Nimmo2017}, open triangles.}
	\label{density}
\end{figure}

It was shown by \citet{Sheppard2008} that larger bodies tend to have larger densities. The authors argue this is due to change of porosity and/or rock/ice ratio. In order to find possible correlation and following \citet{Sheppard2008}, Fig~\ref{density} shows the density estimations (using both our data and values taken from the literature) as a function of absolute magnitude. Only objects larger than $\sim$~200~km (that are considered to be in hydrostatic equilibrium) were used. We found a Pearson correlation coefficient r~=~-0.38, which lies within the 95\% confidence interval and is statistically significant. Thus, we can confirm the existence of a certain correlation.

\section{Conclusions}

We present new photometric observations of nine outer Solar system objects, 6 TNOs and 3 Centaurs. For five objects that were previously observed we combined the published and new data and obtained more accurate rotation periods for three of them. Rotational period value for (281371) 2008 FC76 was reported for the first time. For above-mentioned six objects we were also able to estimate the lower limits of density values. By adding literature densities values to our data set we confirm the existence of a previously reported density/absolute magnitude(or object size) trend.

For three objects which were observed during single nights we were able to estimate a lower limit of the rotational period values and lightcurve amplitudes. We argue the existence of a lightcurve asymmetry for (120178) 2003 OP32 caused by an elongated shape. The rest of the objects exhibit low amplitudes just above the noise level. We were not able to detect any lightcurve asymmetry for them. Nonetheless, we expect, that most of the TNOs and Centaurs population have double-peaked lightcurves caused by (at least slightly) elongated shape. 

\section*{Acknowledgements}

The authors wish to thank A.W. Harris for his constructive review and remarks which helped to improve the paper.

DP has received funding from the European Union's Horizon 2020 research and innovation programme under the Marie Sklodowska-Curie grant agreement n. 664931.




\bibliographystyle{mnras}
\bibliography{photometryTNOs}

\bsp	
\label{lastpage}
\end{document}